\documentclass[5p]{elsarticle}

\usepackage{hyperref}
\usepackage{color}

\journal{Microvascular research}

\begin{document}

\begin{frontmatter}

\title{Spectral analysis of the blood flow in the foot microvascular bed during thermal testing in patients with diabetes mellitus}

\author[mymainaddress]{Irina~Mizeva\corref{mycorrespondingauthor}}
\cortext[mycorrespondingauthor]{Corresponding author}
\ead{mizeva@icmm.ru}
\author[mysecondaryaddress]{Elena~Zharkikh}

\author[mysecondaryaddress]{Victor~Dremin}

\author[mythirdaddress]{Evgeny~Zherebtsov}

\author[mysecondaryaddress]{Irina~Makovik}

\author[mysecondaryaddress]{Elena~Potapova}

\author[mysecondaryaddress]{Andrey~Dunaev}

\address[mymainaddress]{ 614013, Institute of Continuous Media Mechanics,  Korolyov 1, Perm, Russia}
\address[mysecondaryaddress]{302026, Orel State University, 95 Komsomolskaya St, Orel, Russia}
\address[mythirdaddress]{B4 7ET, Aston Institute of Photonic Technologies, Aston University, Aston Triangle, Birmingham, UK}

\begin{abstract}

Timely diagnostics of microcirculatory system abnormalities which are the most severe diabetic complications, is a significant problem of modern health care. Functional abnormalities manifest themselves earlier than the structural one and their assessment is the focus of present-day studies. In this study, the Laser Doppler flowmetry, a noninvasive technique for the cutaneous blood flow monitoring, was used together with local temperature tests and wavelet analysis. The study of the blood flow in the microvascular bed of toes was carried out in control group of 40 subjects, and two diabetic groups differing in the type of diabetes mellitus (17 of DM1 and 23 of DM2).

The temperature tests demonstrated that diabetic patients have impaired vasodilation in response to local heating. The study of oscillating components shows a significant difference of the spectral properties even in the basal conditions. Low frequency pulsations of the blood flow associated with endothelial and  activities are lower in both diabetes groups as well as the ones connected with cardiac activity. Local thermal tests induce variations both in the perfusion and its spectral characteristics, which are different in the groups under consideration. We assume that the results obtained provide a deeper understanding of pathological processes involved in the progress of microvascular abnormalities due to diabetes mellitus.
\end{abstract}


\end{frontmatter}

\section{Introduction}
\label{intro}

In recent years, diagnosis, care and treatment of patients with diabetes mellitus (DM) have been the highest healthcare priorities. In 2016, over 415 million people worldwide were diagnosed with diabetes (estimates from the International Diabetes Federation.)\cite{IDF}. This number is expected to increase to 642 million people by 2040. Clinical observations demonstrate that persistently high blood sugar can damage blood vessels and nerves and that microvascular abnormalities may appear already in the preclinical phases of diabetes \citep{Caballero1999,Smirnova2013}.

Microcirculation disorders manifest themselves in all parts of the body and affect the functioning of various organs, including kidneys, eyes, cardiovascular system and skin. This significantly reduces the life quality of patients and may lead to their full disability.

Diabetic foot ulcer is a major DM complication, including permanent disability and even amputations at a late stage \cite{Fuchs2017}. Timely diagnosis, monitoring and treatment of the complications reduce the severity of their manifestation and  potentially prevent their further development \cite{Schramm2006}.

Assessment of the microcirculation may conveniently be performed in skin because of its ease accessibility. The  cutaneous blood flow can be evaluated using various optical diagnostic methods \cite{Daly2013}, of which Laser speckle, videocapillaroscopy, optical coherence tomography, and laser Doppler flowmetry (LDF) are most frequently used.

LDF \cite{Stern1975}  allows one to estimate the blood flow in the microvasculature  \textit{in-vivo}. It is based on optical non-invasive sensing of tissue using laser light and further analysis of the scattered radiation partially reflected by the moving red blood cells. Spectral analysis of the LDF signal is widely used to assess properties of the microcirculation system \cite{Stefanovska}. There are five frequency bands corresponding to different regulatory mechanisms. The pulse (0.45-1.6 Hz) and respiratory (0.2-0.45 Hz) bands carry information about the influence of heart rate and movement of the thorax on the peripheral blood flow.  The myogenic mechanism of vascular tone regulation mirrors the response of vascular smooth muscle cells to the transmural pressure. Blood flow oscillations at frequencies  (0.05-0.15 Hz) characterize  its activity. Neurogenic sympathetic vasomotor activity induces vessels walls movement with frequency 0.02-0.05 Hz. Slow waves in blood flow  (0.0095-0.02 Hz) reflect the vascular tone regulation due to the endothelium activity. The borders of these frequency bands slightly vary from paper to paper \cite{Stefanovska,Hodges2017}, additional frequency bands are allocated\cite{Krupatkin2007}. Regional differences in the cutaneous microvascular function should be taken into account when analyzing variations in skin blood flow \cite{Hodges2014, Sorelli2017}. The microcirculatory dysfunction in diabetes usually manifest itself in the feet, that is why we examine them in this work.

Functional microcirculation abnormalities  appear earlier than the structural ones \cite{Beer2008}. The difference between normal and pathological conditions in LDF records in basal conditions wasn't revealed \cite{Walther2015}. It is attributed to physiological variations in skin blood flow and limitations of the LDF technique \cite{Fredriksson2007, Mizeva2016, JBPE3129}. A promising method for the microvasculature functional state monitoring  is based on the estimation of dynamic variations in cutaneous blood flow \cite{Geyer2004,Humeau2004}. The functioning of the microvasculature is often  evaluated by analyzing the impact of stress tests: thermal, mental, pharmacological, orthostatic, breath and occlusive. Being noninvasive and easy to implement, thermal tests are most widespread \cite{Roberts2017}. Diabetes primarily damages unmyelinated nociceptive C-fibers, which are activated by heating above $42^{\circ}$C \cite{Campero2009}. That's why the heating tests are implemented to assess microvascular abnormalities in subjects with diabetes \cite{Jan2013,Parshakov2017}.

Both reflexes vasodilation and vasoconstriction mirror the function of blood flow regulative mechanisms \cite{Sheppard2011}. Managed by sympathetic vasoconstrictor nerves \cite{Pergola1993} microvacular system is able to decrease blood flow at low temperature. On the other hand, change in the responsiveness of smooth vascular muscles to sympathetic stimulation during local cooling \cite{Stephens2001} causes vasoconstiction which is  impaired in patients with DM \cite{Sivitz2007}.

Slow heating ($0.5^0$ C per 5 min or more) \cite{Hodges2009} reduces the effect of microcirculatory system response, while  rapid heating ($0.5^{\circ}$ C per 5 s or less ) up to a temperature above $39^{\circ}$ C shows the pronounced reproducible increase of perfusion \cite{Roberts2017, Dremin2016}. In this work, local thermal tests at different temperature ($25 ^{\circ}$C, $35 ^{\circ}$C and $42 ^{\circ}$C) are performed consequently. We assume that this sequence of local thermal stimuli promotes pronounced activation of the local regulatory mechanisms of blood flow. In particular, rapid local heating up to $35^{\circ}$C should induce an axon-reflex \cite{Johnson2014} due to  the activation of sensory peptidergic nerve fibres \cite{Stephens2001}. Further heating up to $42^{\circ}$C provoked the development of vasodilation associated with the release of nitric oxide (NO) from the vascular endothelium \cite{Minson2001}. In the work, all the suppositions are well consistent with the obtained experimental results.

The aim of this study is to analyze the variations in the main microhaemodynamic parameters in different modes of heating on the feet of patients with DM1 and DM2.

\section{Materials and methods}

\subsection{Groups of subjects}

The study involved 40 patients from the Endocrinology Department of the Orel Regional Clinical Hospital (Russia) with DM1 and DM2 (the main characteristics of patients are shown in Table \ref{Tab1}). The exclusion criteria were the presence of foot ulceration at the time of recruitment, end-stage renal disease (renal dialysis or kidney transplantation), and any other serious chronic disease that can affect wound healing. The clinical and laboratory parameters were measured by standard laboratory procedures. Blood pressure measurement was carried out after 5 min of the patient rest in a sitting position. Mean age of this group was $43\pm10$ years, and it includes patients with both types DM1 and DM2 (Table~\ref{Tab1}). These two groups had slightly different mean age, but close the disease duration.

\begin{table}[h]
\caption{\label{jlab1} The main characteristics of the groups under study}
\begin{tabular}{c  c c c c }
\\\hline
& controls &DM1 &DM2 &Reference values of the laboratory \\
&n=40 & n=17 & n=23 \\\hline
Sex (M/F) & 26/14 & 10/7 & 10/13 \\
Foot temperature,  $^{\circ} C$ & $27\pm1$ & $30\pm1$ & $30\pm2$ \\
Age, years & $27\pm1$ & $35\pm9$ & $50\pm6$ \\
Diabetes duration, years &  $39\pm9$ & $14\pm10$ & $7\pm6$ \\
Body mass index, $kg/m^2$ &  & $25\pm5$ & $35\pm5$ \\
Fasting glucose, $mmol/l$ &  & $8.1\pm4.7$ & $9.2\pm3.5$ \\
HbA1c, \% & & $7.9\pm0.8$ & $8.8\pm0.9$ & $4.0-6.0$ \\
Total cholesterol, $mmol/l$ &  & $4.6\pm0.9$ & $5.4\pm0.9$ & $3.5-5.0$\\
Creatinine, $umol/l$ &    & $88\pm37$ & $74\pm16$ & $70-110$ \\
Urea, $mmol/l$ &  & $6.1\pm3.4$ & $5.7\pm1.8$ & $2.5-7.5$ \\
ALT, $IU/L$ & & $27\pm14$ & $36\pm18$ & $10-38$ \\
AST, $IU/L$ & & $26\pm10$ & $27\pm10$ & $10-40$ \\
Systolic BP, $mmHg$ & & $124\pm16$ & $141\pm13$ \\
Diastolic BP, $mmHg$ & & $78\pm7$ & $86\pm6$
\\ \hline
\end{tabular}
\label{Tab1}
\end{table}

The control group consisted of 40 healthy volunteers (26 male, 14 female) with a mean age of $39\pm9$ years and no diagnosed diseases of the circulatory system, musculoskeletal system or connective tissue. The volunteers with exacerbations in diseases of the cardiovascular, pulmonary, neuroendocrine system, gastrointestinal tract, liver, kidneys, blood and any other serious chronic diseases, which could influence  the microcirculation system, were excluded from the study as well as subjects with an alcohol history, medication or drug abuse. The study protocol was approved by the ethics committee at the Orel State University named after I.S. Turgenev. After receiving the description of the protocol, all  volunteers signed the informed consent form.

\subsection{Samples collection method}

\begin{figure}
\centering
\resizebox{0.8\columnwidth}{!}{\includegraphics{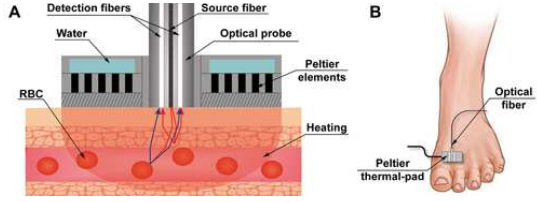}}
\caption{Coupling of the optical probe with the Peltier element (a); localization of the probe on a human lower limb (b)}
\label{fig:1}
\centering
\end{figure}

The cutaneous blood flow was examined using the laser Doppler flowmeter (SPE “LAZMA” Ltd., Russia). The Doppler channel is built by using single mode laser module with a wavelength of 1064 nm. A fiber optical probe was used to deliver laser light and register the reflected, shifted in frequency,  radiation from the tissue. Probe radiation power at the output of the fiber probe does not exceed 1.5 mW. The geometric parameters of the fiber  probe (light emitting fiber 6 $\mu m $ in diameter,  light-collecting fiber 400 $\mu m$ in diameter, probes were separated in 1.5 mm) give the receiving numerical aperture 0.22. The average diagnostic volume of the probe is 1-3 $mm^3$ (Fig.~\ref{fig:1}a).

The optical probe was installed into the hole of the Peltier element temperature controlled by a thermistor with the accuracy of $0.1^{\circ}$ C. This system (Peltier element together with the optical probe) was mounted on the dorsal surface of the foot at a point located on the plateau between the 1st and 2nd metatarsals (Fig.~\ref{fig:1}b).

Before the measurement, carried over 2 hours after a meal, volunteers were adapted for at least 10 min at room temperature.  All studies were performed in subjects lied in the supine position. In the basal conditions, subjects had different skin temperatures (Table \ref{Tab1}). To unify  measurements, we have chosen the cooling to $25^{\circ} C$. Each study included four steps: the basic test for 4 min, cooling to $25^{\circ} C$ for 4 min, and local thermal tests at temperatures of $35^{\circ} C$ and $42^{\circ} C$ for 4 and 10 min, respectively. Thus the duration of measurement of one foot was 22 minutes, consequently both feet were investigated. The LDF sampling on each leg was collected continuously. In this work,  we discuss the results obtained from one leg to exclude long staying in the supine position.

\subsubsection{Data preprocessing and analysis}
\label{sec:DataAnalysis}

The LDF signal was decomposed using a wavelet transform as:
\begin{equation}
W(\nu,\tau)=\nu\int\limits_{-\infty}^{\infty}f(t)\psi^{*}(\nu(t-\tau))dt,
\label{CWT}
\end{equation}
where $^*$ means complex conjugation. The Morlet wavelet written in the form
\begin{equation}
\psi(t)=e^{2\pi i t}e^{-t^2/\sigma} \label{morlet}
\end{equation}
was used for the series expansion in the decay parameter $\sigma=1$. Integrating the power over time gives the global wavelet spectrum
\begin{equation}
M(\nu)=\frac{1}{T}\int_0^T |W(\nu,t)|^2 dt.  \label{eq:spec}
\end{equation}

The wavelet spectrum is a smoothed version of the Fourier spectrum \cite{Frick2015}. We calculated wavelet coefficients for the frequency range 0.01-2 Hz with the logarithmic partitioning on 50 frequency bands. At the first step we calculated $M(\nu)$ for every record and departed from boundaries and LDF stepwise variations caused by changes of environmental conditions to exclude their influence on the spectrum. The integral wavelet spectra were averaged over the group. For each frequency band we obtained energy distribution, which was compared for health and pathological groups. The frequency bands corresponding to different physiological mechanism are shown on the plots for the reference.

The Mann-Whitney test was used to compare the intergroup results and the Wilcoxon statistical test for intragroup variations. All data processing was implemented using Mathmatica 8.0, Wolfram Research.

\section{Results}

\begin{figure}
\begin{center}
\resizebox{0.8\columnwidth}{!}{\includegraphics{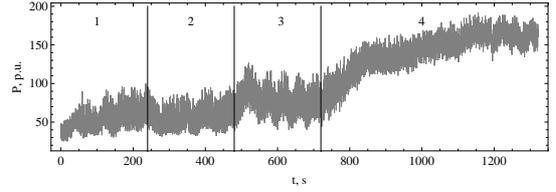}}
\caption{Typical LDF sample collected from patient with diabetes (from the right foot, diabetes duration 30 years). Numbers indicate experimental stages: 1 - record in basal conditions, 2 - record during cooling,  3 - first heating up to $35^{\circ} C$, 4 - second heating up to $42^{\circ}C $ }
\label{fig:2}
\end{center}
\end{figure}

\begin{figure}
\begin{center}
\resizebox{0.8\columnwidth}{!}{\includegraphics{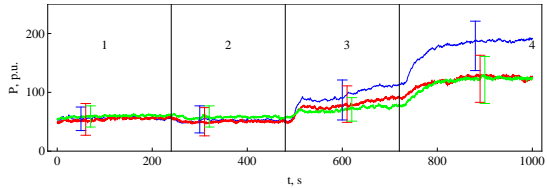}}
\caption{Dynamics of perfusion averaged over all measurements( blue - controls, red - patients with DM1, green - patients with DM2). First, we applied the moving average filter with a window of 0.25 s and then estimated mean value at each instant.  Error bars indicate a mean standard deviation at a certain stage of the experiment. Numbers show experimental stages:  1 - basal conditions, 2 -  cooling,  3 - first heating up to $35^{\circ} C$, 4 - second heating (up to $42^{\circ}C $) }
\label{fig:3}
\end{center}
\end{figure}

\begin{figure}
\begin{center}
\resizebox{0.8\columnwidth}{!}{\includegraphics{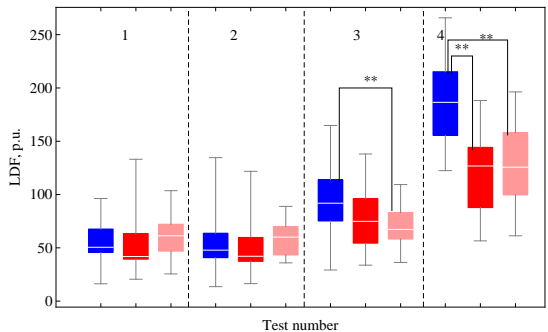}}
\caption{Box-Whisker diagram of mean perfusion during 4 experimental stages 1 -  basal conditions, 2 - cooling, 3 - $35^{\circ}$ C heating, 4 -  $42^{\circ}$ C heating. Blue rectangles correspond to the control group, red – patients with DM1, light red - DM2. By stars we mark the level of significance estimated using Mann-Whitney test ( ** - $p<0.01$). Increasing of the perfusion during 3 and 4 tests is also significant. In the both groups tests provoke significant variation of perfusion (the level of significance was estimated using Wilcoxon test, $p<0.001$)}
\label{fig:4}
\end{center}
\end{figure}

An example of the collected $LDF$ samples is presented in Fig.~\ref{fig:2}. Thermal tests  provoke significant variations both in the average perfusion \ref{fig:4}  and its oscillation component (see for example the third experimental stage in Fig.~\ref{fig:2}). The statistical  analysis (Fig.~\ref{fig:3}) demonstrates close values of $P$ without any significant differences in the basal conditions. Cooling provokes weak vasoconstriction, and heating -- significant vasodilation. Note that the vasodilation dynamics varies in groups (Fig.~\ref{fig:3}). All measurements showed  a peak at the beginning of  heating caused by the axon-reflex. The highest rate of vasodilation is found in the control group and this characteristic is impaired in both diabetic groups.   Moreover, the  heating clarified the difference between groups, and perfusion of the heated skin significantly differs in the groups of healthy and diabetes subjects, but is similar in two diabetes groups.

\subsection{Basal conditions}

\begin{figure}
\begin{center}
\resizebox{0.48\columnwidth}{!}{\includegraphics{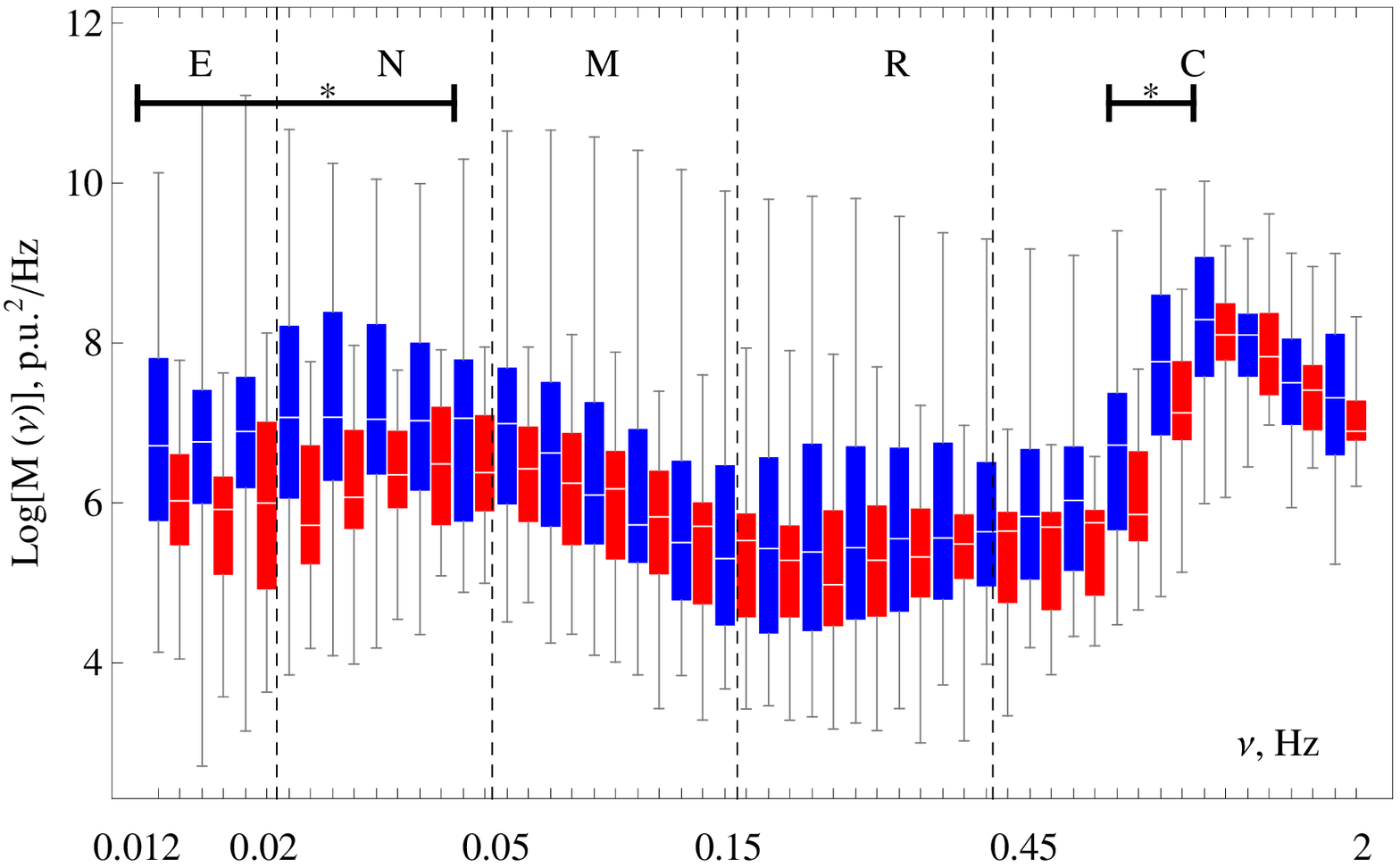}}
\resizebox{0.48\columnwidth}{!}{\includegraphics{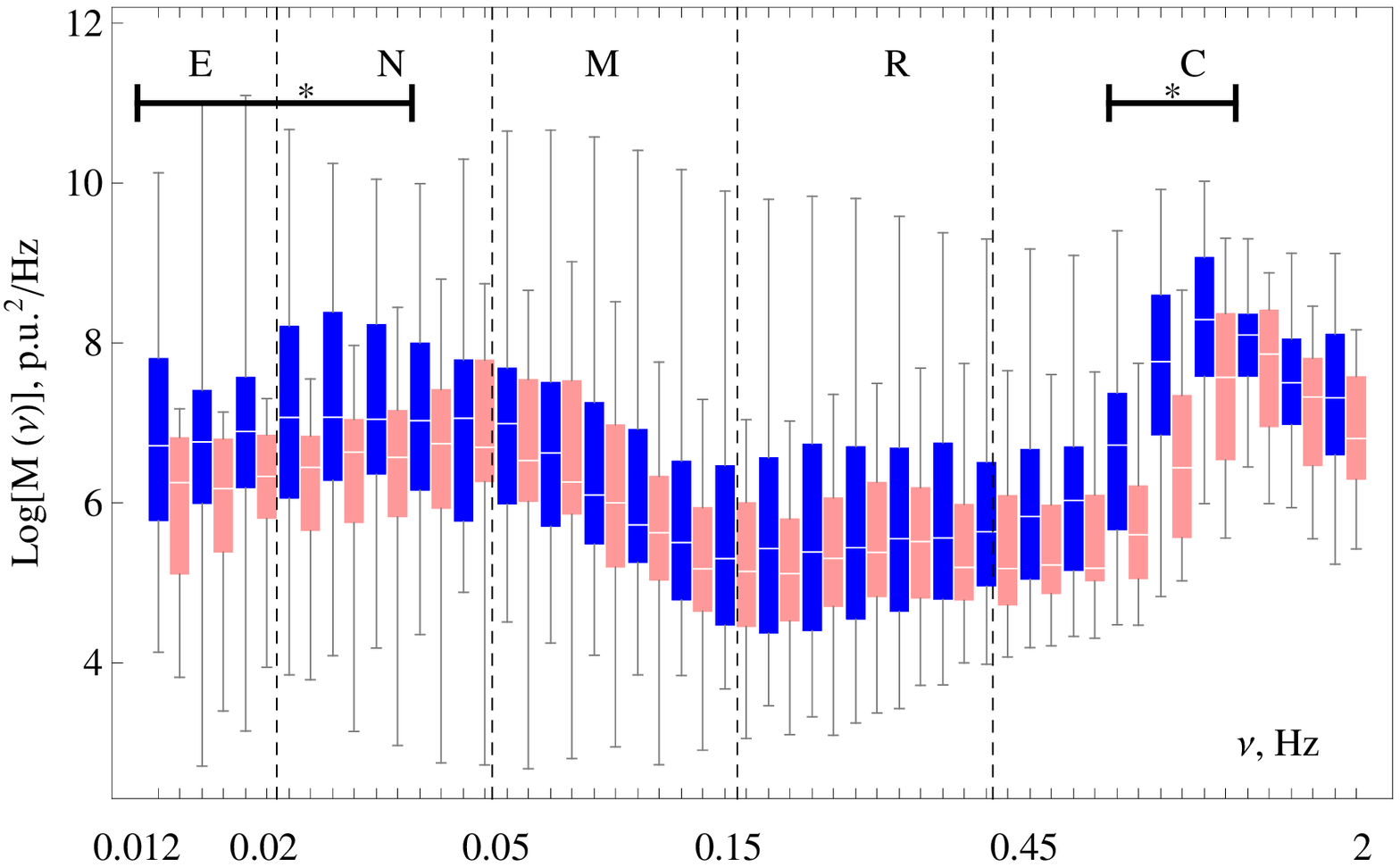}}
\caption{Averaged spectra of LDF samples in basal conditions:  blue rectangles - control group. The bright red boxes (left panel) indicates $M(\nu)$ for the group of DM1 diabetic patients and the light red colour (right panel) - $M(\nu)$ for the group of DM2. Thick lines in the upper parts of the plot indicate frequency band where $M(\nu)$ significantly differs ($p<0.05$) }
\label{fig:5}
\end{center}
\end{figure}

The lowest perfusion in the basal conditions  was observed in the control group ($P=53\pm18$ p.u.). Patients with both types of diabetes had slightly higher perfusion, $54\pm27$ p.u. in DM1 and $58\pm20$ p.u. in DM2. Both diabetes groups had impaired amplitude of perfusion oscillation in the frequency range (0.012-0.045) Hz (Fig.~\ref{fig:5}) in comparison with the control group. These frequencies fall into the intervals, which correspond to neurogenic and endothelial vascular tone regulation mechanisms. Fluctuations in the range 0.5-1 Hz were also weaker in both diabetes groups. Moreover, oscillations of these frequencies were significantly lower in patients with DM2 than in patients with DM1.

\subsection{Local cooling}

\begin{figure}
\begin{center}
\resizebox{0.48\columnwidth}{!}{\includegraphics{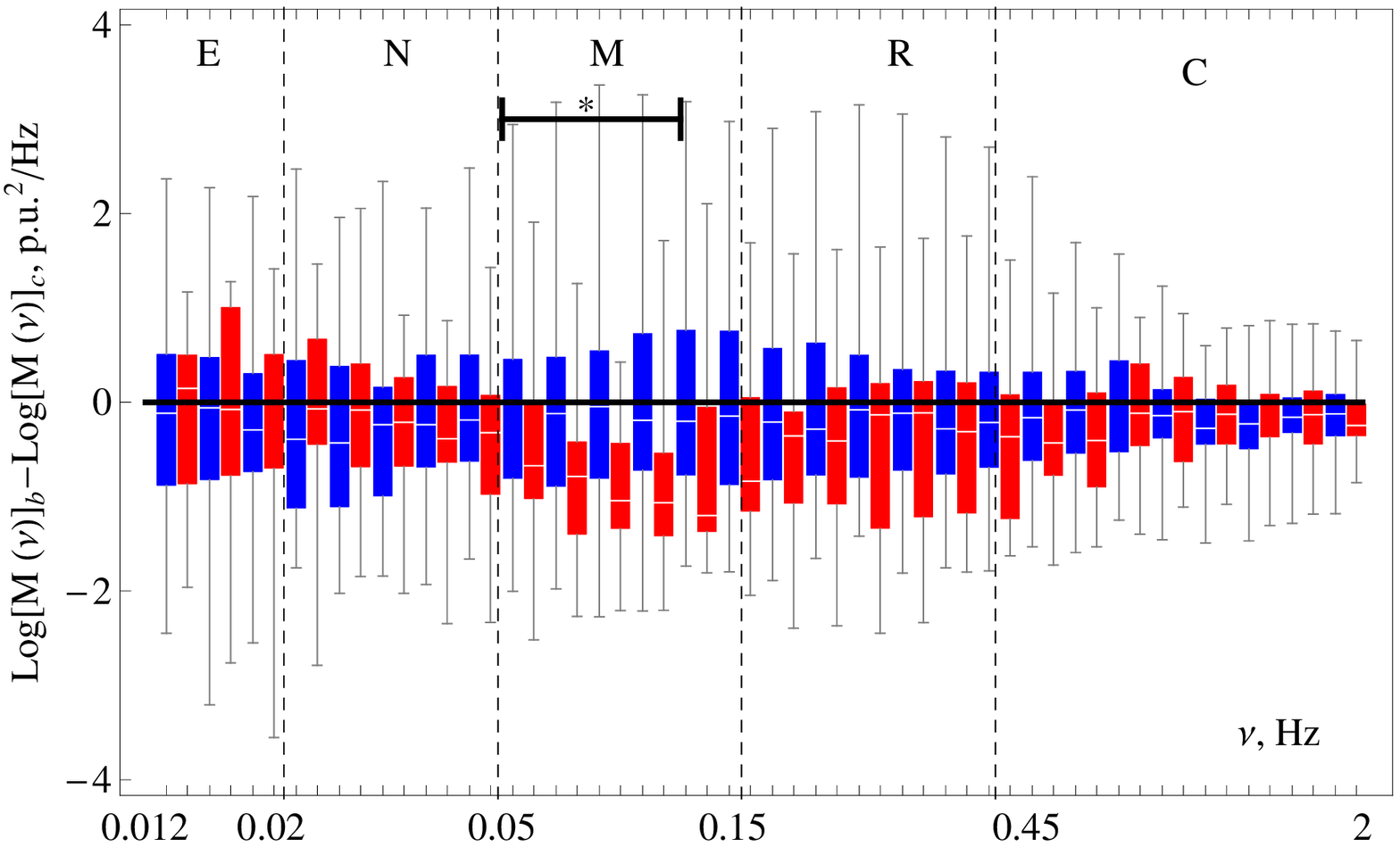}}
\resizebox{0.48\columnwidth}{!}{\includegraphics{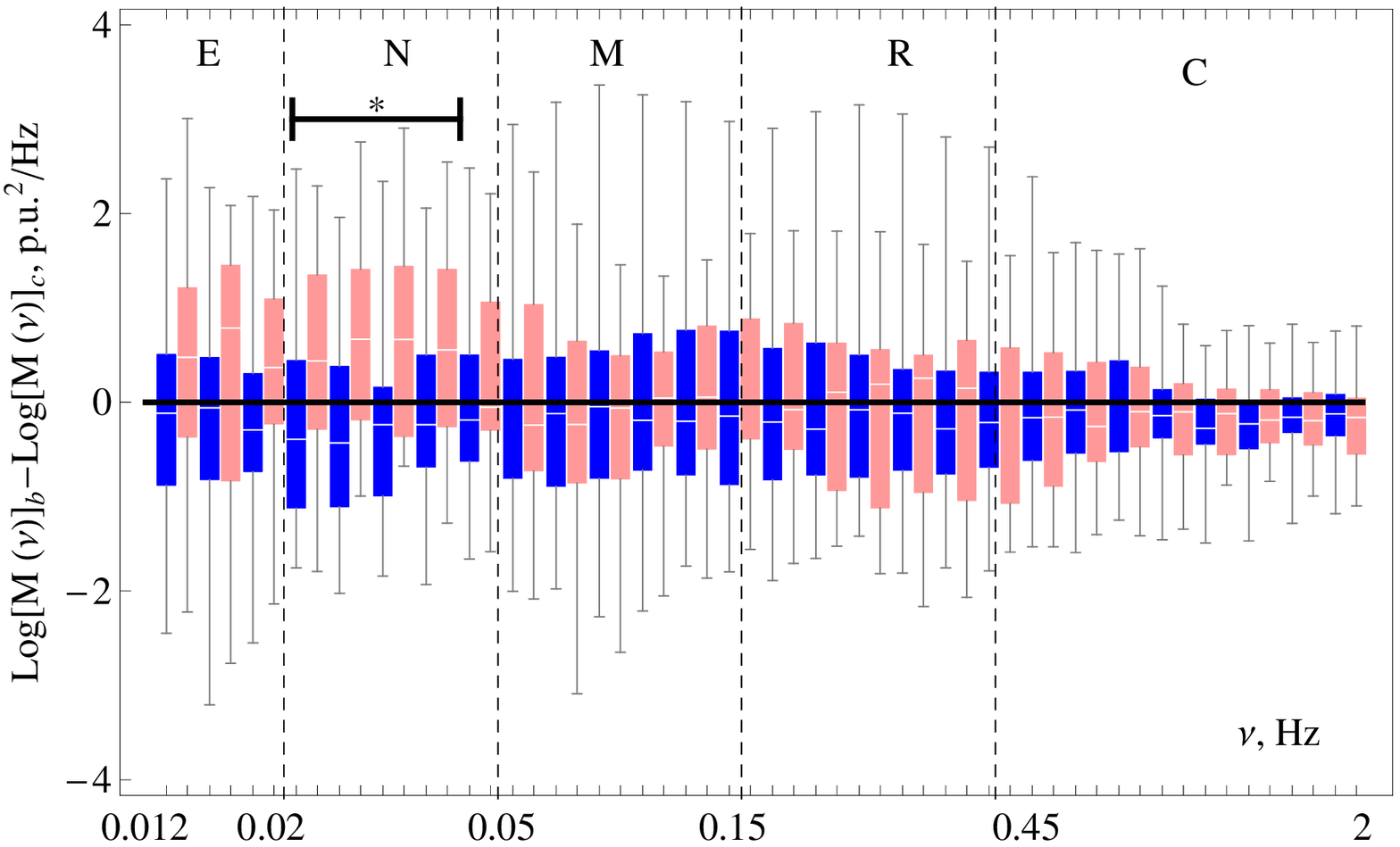}}
\caption{Variation of the spectral energy caused by the cooling. For every frequency we calculate $M(\nu)_c$ during cooling ($M(\nu)_c$) and in basal conditions ($M(\nu)_b$), then estimate their difference for every frequency $M(\nu)_c-M(\nu)_b$ for every LDF sample and then construct the  Box Whisker diagram. By thick lines in the upper part of the plot frequency bands where the variation of energy of pulsation is significant ($p<0.05$) are marked. By blue rectangles on both plots we demonstrate results obtained for the control group and by red boxes - for diabetic patients of type 1 on the left panel and type 2 on the right.}
\label{fig:6}
\end{center}
\end{figure}

Local cooling-induced vasoconstriction causes the variation in the spectral properties of LDF signals. The averaged perfusion is $46\pm16$ p.u. in the control group, $50\pm23$ p.u. in the DM1 group, and $55\pm15$ p.u. in DM2. To study the phenomenon of spectral variation, we estimated the difference between $M(\nu)$ in basal conditions and at cooling for every subject. The results obtained are presented as a box-whisker plot (Fig.~\ref{fig:6}). The spectral characteristic of LDF samples of healthy subjects only slightly varies in response to cooling.

At exposure temperature of $25^{\circ}$C  a significant reduction in oscillations was observed  in the frequency band of 0.05-0.14 Hz in patients with DM1. This frequency band falls within the range associated with the myogenic activity. At the same time, spectral energy of the LDF signal of DM2 patients in this frequency band was constant, but the oscillations of 0.02-0.04 Hz characterising neurogenic activity significantly increased during cooling. Note that the amplitude of pulsations between two diabetic groups was markedly different in these two frequency bands.

\subsection{Local heating ($35^{\circ}$C)}

Local heating up to $35^{\circ}$C provoked vasodilation. the difference of averaged perfusion became significant between the examined groups, perfusion increased to $92\pm 28$ p.u. in healthy subjects and still remained slightly lower at the level of $79 \pm 30$ p.u. in DM1 and $67 \pm 16$ p.u. in DM2.

\begin{figure}
\begin{center}
\resizebox{0.48\columnwidth}{!}{\includegraphics{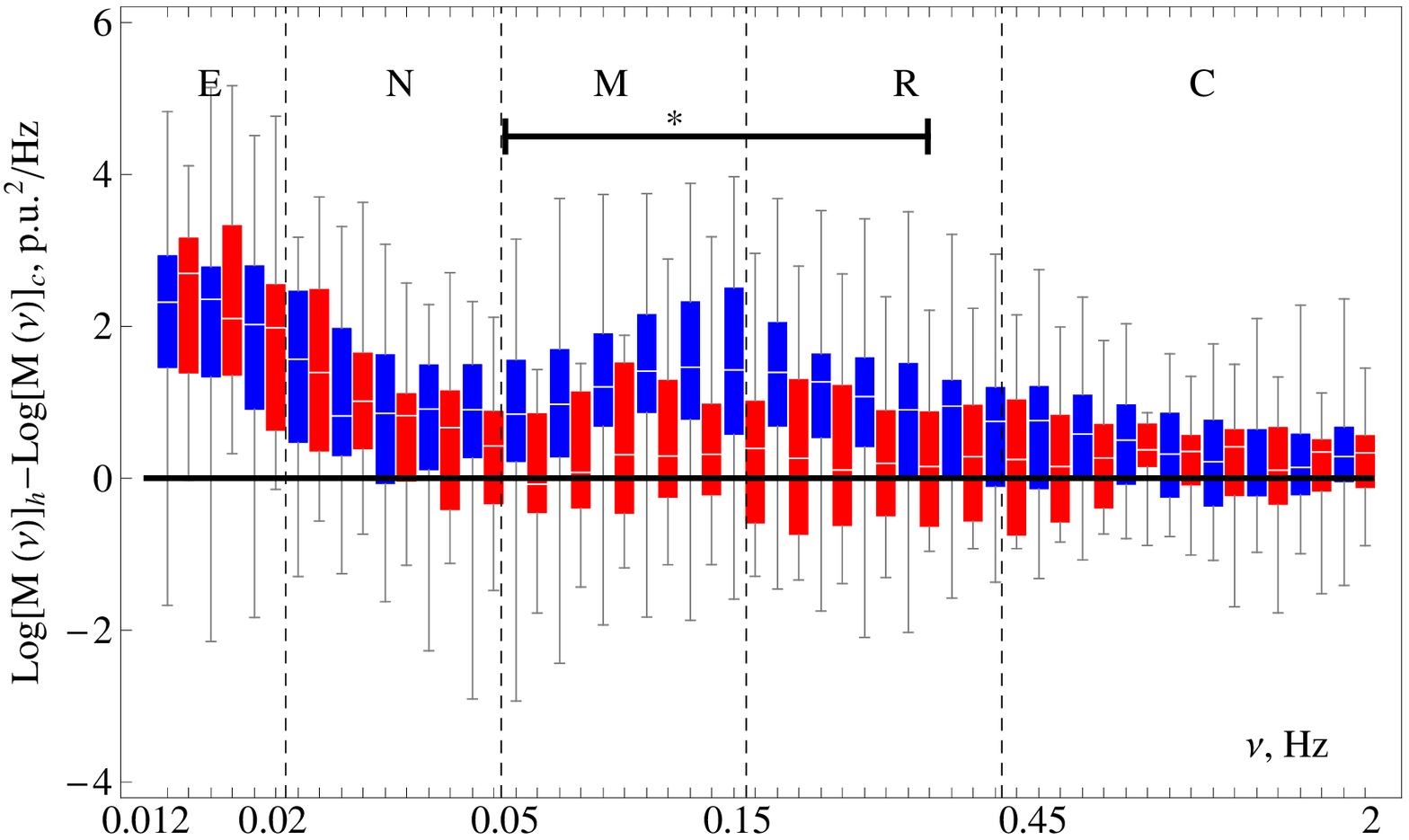}}
\resizebox{0.48\columnwidth}{!}{\includegraphics{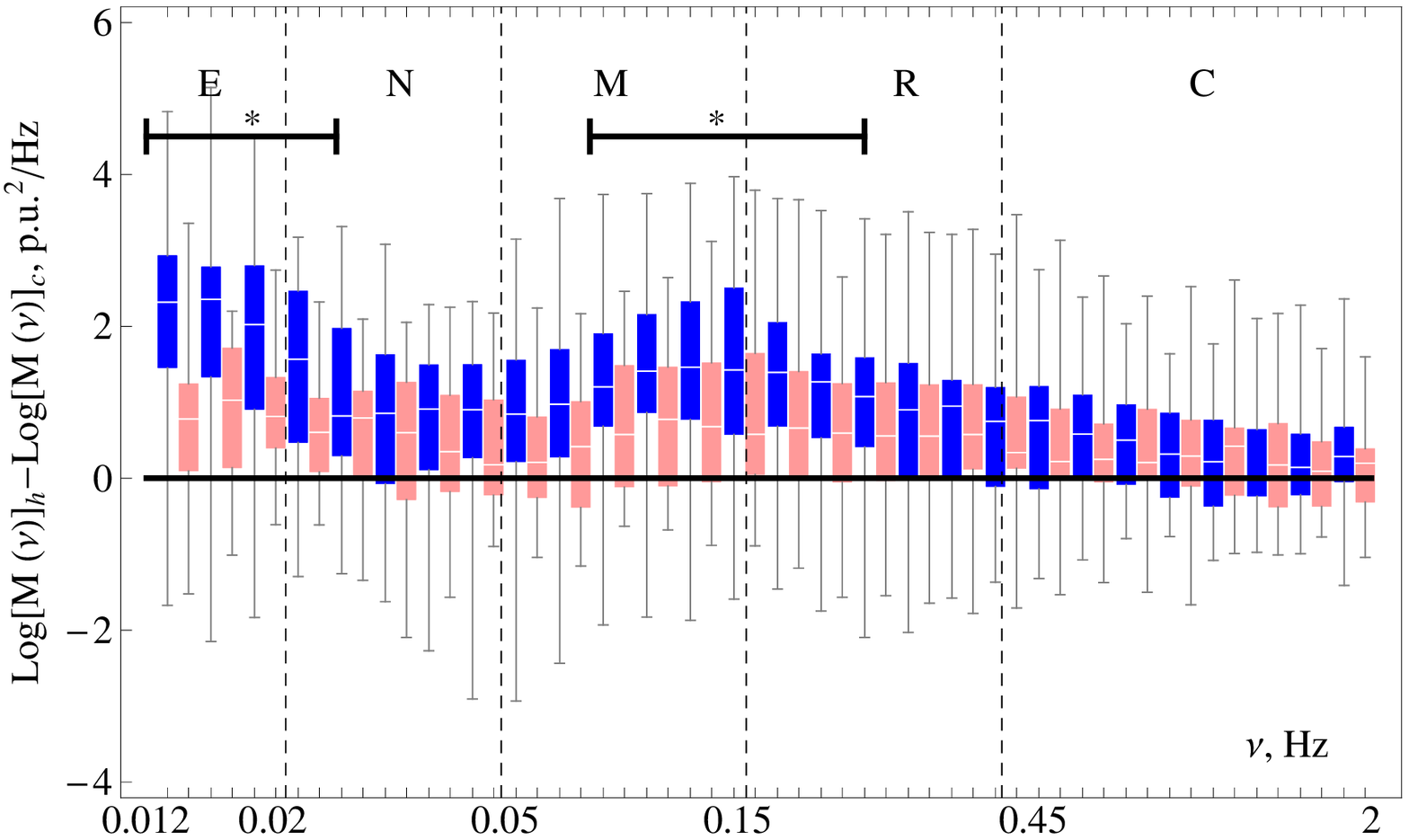}}
\caption{Variation of the spectral energy ($M(\nu)_h-M(\nu)_b$)
caused by the heating up to 35 C. The algorithm of the plot construction is similar to one applied in the Fig.\ref{fig:6}. Blue boxes result for the control group, red - diabetic patients (type 1 on the left panel, type 2 - on the right.)}
\label{fig:7}
\end{center}
\end{figure}

At this stage, oscillations in the frequency band 0.05-0.45 Hz increased in the control and both patient groups (Fig.~\ref{fig:7}). This confirms that a rapid temperature rise can stimulate nociceptive receptors and increase the neurogenic component of the signal \cite{Roberts2017}. There is a sharp peak for the controls (in  Fig.\ref{fig:7}) at the frequency close to 0.14 Hz. The variation mentioned above is significantly lower in both patient groups in comparison with the control group. The smallest reaction was observed in DM1 subjects. Together with oscillations corresponding to the myogenic activity, low frequency oscillations increased in all three groups. The difference in the amplitude of oscillations in basal and heated states was found to be  significant for all three groups. Taking into account the relation between pulsatile and averaged components of LDF signal \cite{Mizeva2016} and rising the average perfusion due to heating, it is difficult to explain this result. On the other hand,  averaged perfusion in the groups with diabetes is statistically equal ($p>0.05$), but the amplitude of oscillations in the endothelial frequency band differs significantly. Hence, we can conclude that the mechanisms involved in the low frequency modulation of cutaneous blood flow are strongly related to DM1.

\subsection{Prolongated local heating ($42^{\circ}$C)}

\begin{figure}
\begin{center}
\resizebox{0.48\columnwidth}{!}{\includegraphics{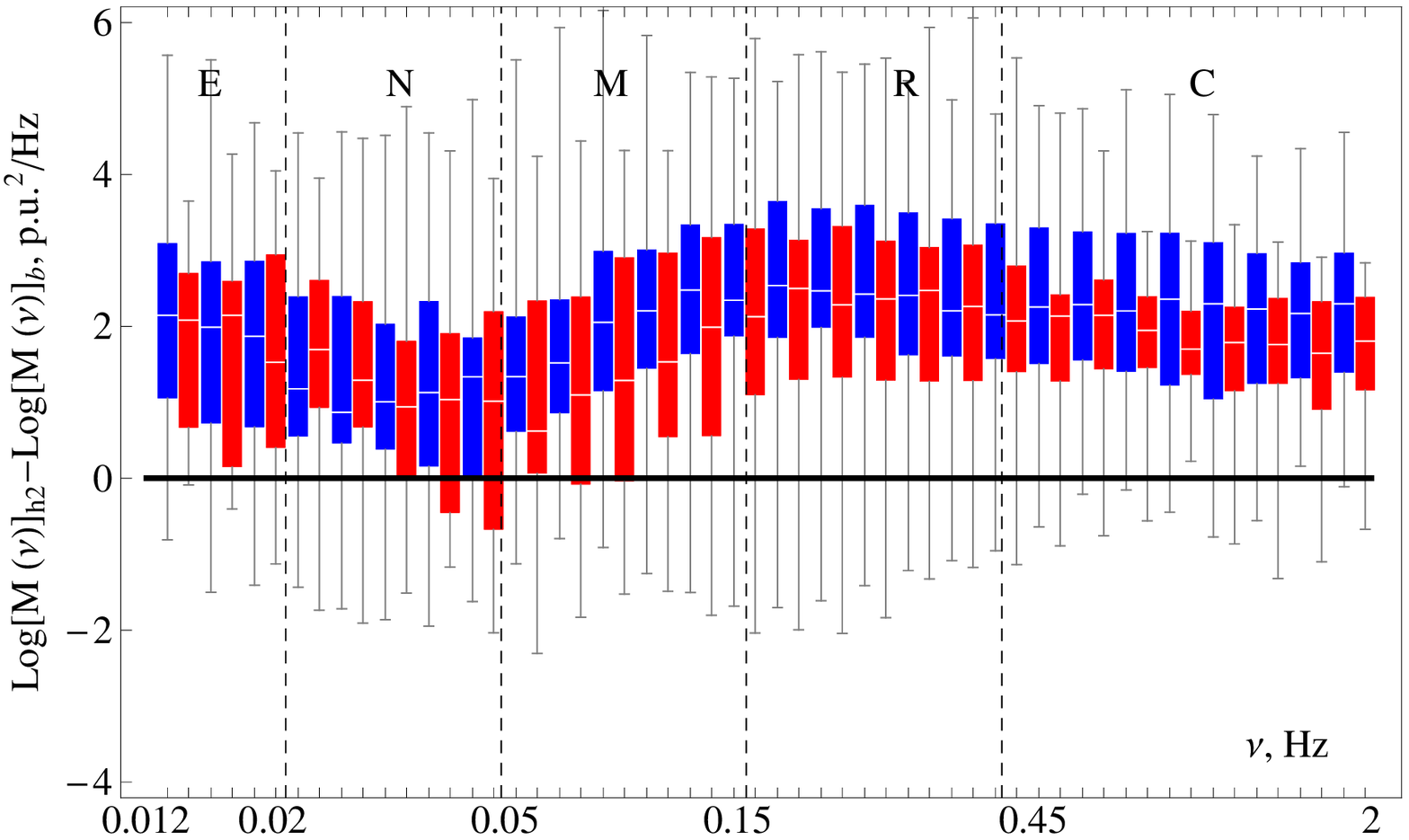}}
\resizebox{0.48\columnwidth}{!}{\includegraphics{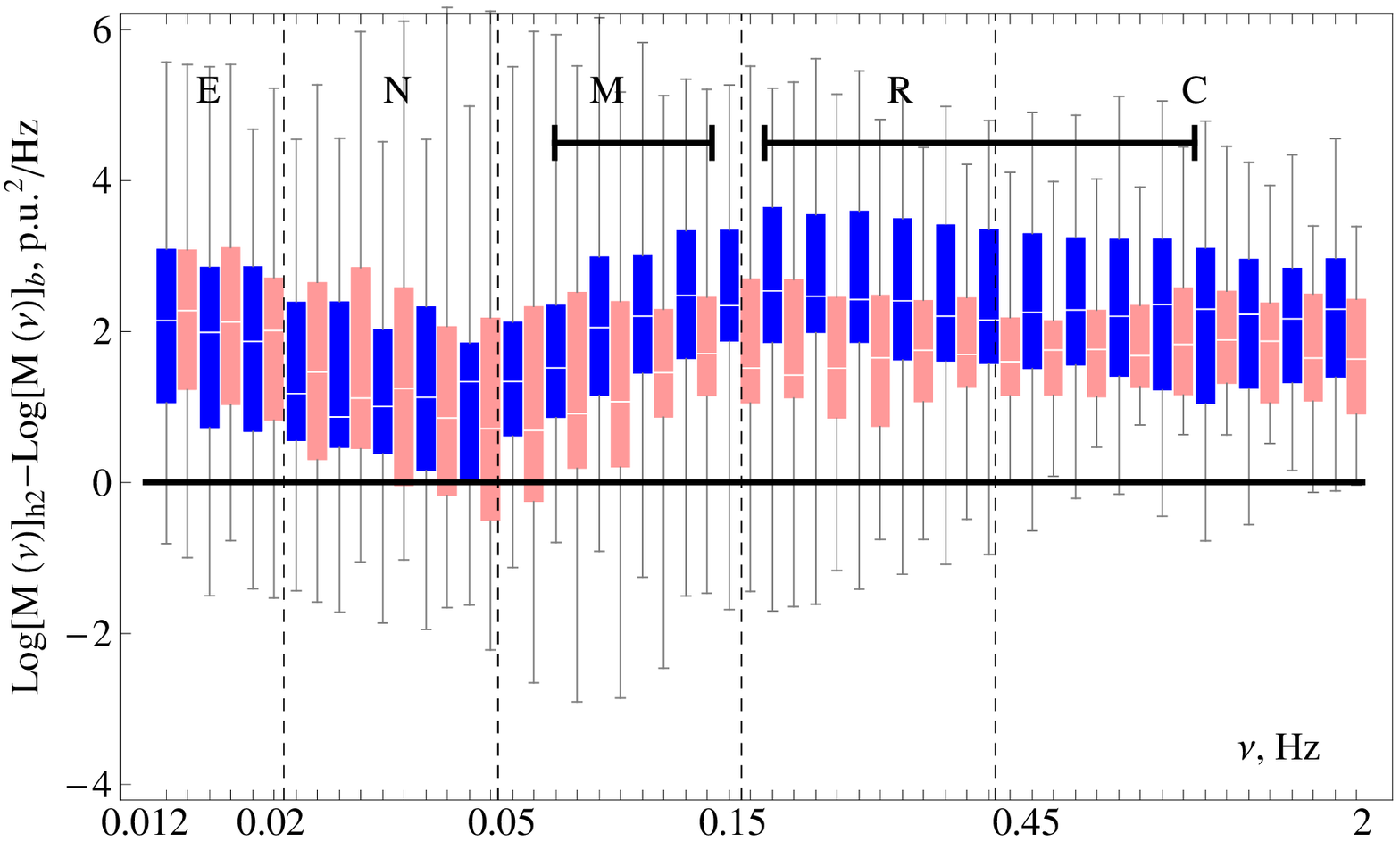}}
\caption{Variation of the spectral energy ($M(\nu)_h-M(\nu)_b$)
caused by the heating up to 42 C. The algorithm of the plot construction is similar to one applied in the Fig.\ref{fig:6}. Blue boxes results for the control group, red - diabetic patients (type 1 on the left panel, type 2 - on the right.) }
\label{fig:8}
\end{center}
\end{figure}

The next stage of the experiment was prolongated heating up to $42^{\circ}$ C. Higher temperature provokes stronger vasodilation, the mean $P$ rose up to $190 \pm 27$ p.u in the control group. Vasodilation response was impaired in both patients groups  ($128\pm38$ p.u. in DM1 and $122\pm38$ p.u. in DM2). Local heating up to $42^{\circ}$C caused an increase in all frequency bands (Fig.~\ref{fig:8}). Note that the amplitude of oscillation  is lower in both diabetes groups in comparison with controls, and it is significant for high frequency pulsations of patients with DM2.

\section{Discussion}

In this paper we have analyzed  variations in the  cutaneous blood flow  caused by local cooling and heating measured using LDF. In the basal conditions  all subjects had a similar level of perfusion, which is slightly higher in DM2 patients, which is fully coincide with \cite{Schramm2006,Burns2012}.

Even in rest conditions, the analysis of blood flow oscillations reveals a significant difference in  the microhaemodynamic parameters of healthy and pathological subjects. We observed lower amplitude of 1 Hz oscillations in both DM groups  in comparison with controls in basal conditions. As local microvascular tone regulation mechanisms are not involved in modulation of the cardiac activity we assume that this difference is associated with morphological abnormalities of microvascular system in diabetes. The lowest energy of 1 Hz pulsations was observed in patients with DM2, in DM1 energy was slightly higher. However both these values were significantly lower than in the control group.  As the cardiac stroke volume \cite{Devereux2000} in patients with DM is higher than in healthy ones we conclude, that cardiac wave is stronger dumped in DM patients then in healthy ones by cardivascular system. Metabolic syndrome, insulin resistance, impaired glucose tolerance and accumulation of advanced glycation end products are positively correlated with increasing arterial stiffness \cite{Zieman2005}, thus we assume that the difference in the blood flow oscillations associated with cardiac activity indirectly characterizes vessels elastic properties and indicate rising arterial stiffness \cite{Jaiswal2013} of diabetic patients.

Autonomic neuropathy,  correlated with the development of endothelial dysfunction and increasing of the arterial stiffness in patients with DM, leads to impaired low frequency vasomotions (0.01-0.05 Hz)  established in this paper.

Smooth vascular muscles response to sympathetic system stimulation during local cooling \cite{Stephens2001} provokes vasoconstriction. The spectral characteristics of controls' LDF signals practically  are not disturbed by cooling; small variations are observed on the left end of the neurogenic frequency band. Patients with DM1 demonstrated decrease in myogenic oscillations caused by local cooling. Another response demonstrate patients with DM2, namely, the neurogenic component of vascular tone regulation increased, the difference became significant.

Local mild heating initiates the sequence of reflexes, which leads to vasodilation \cite{Johnson2010}. At the beginning of heating we observed a local peak on the perfusion-time curve associated with the axon reflex; vasodilation and its rate were impaired in DM patients in comparison with controls.Further, after the local minimum on the perfusion-time curve vasodilation is supplied by inducing NO \cite{Johnson2010}. Perfusion on this stage was impaired in DM patients similar to \cite{Stevens1995, Stevens1994}. Having the highest perfusion in basal conditions, the patients with DM1 have the lowest one under local heating conditions. In \cite{Mizeva2017} such behavior  was interpreted as a low reserve of the microcirculation system in pathological conditions. The LDF record during this test is nonstationary and the its slow variations are related with the axon-reflex peak in the first part of the test and to the endothelial activity in the second. For this reason the discussion of the low-frequency part of the spectra is dropped from the consideration. We revealed the increase of myogenic activity in all groups, its highest variation is in the control group is accompanied by a  sharp peak at the frequency close to 0.14 Hz. Similar behavior in \cite{Sheppard2011} was associated  with high pre-capillary pressure and stretching the arterioles causing myogenic oscillations. DM patients have lower responce in myogenic activity on the heating.

Prolongated heating induces NO mediated vasodilation, which is lower values in DM patients we associate it with the endothelial dysfunction \cite{Shi2017}. Important factors for endothelial dysfunction are decreased production and bioavailability of NO, increased production of vasoconstrictors (endothelin-1), oxidative stress and increasing angiogenesis which are typical for diabetes. Prolongated heating test induces nonuniform increase in perfusion oscillations. The spectra of the perfusion-time curve for healthy volunteers had the lowest impact of neurogenic activity associated pulsations comparing to another frequency bands. It indicates the key role of the endothelial activity in the vasodilation on this stage. Both types of DM impair vasomotions and  DM2 gives the reliable difference. Further \cite{Mizeva2017} we do not mention the reliable difference in 0.01-0.02 Hz oscillations in healthy and pathological conditions.

In this paper we propose a modified version \cite{Mizeva2017JBO} of the commonly used  approach to analyze the spectral properties of LDF signals. Frequently \cite{Stefanovska,Krupatkin2007} the wavelet spectra of LDF signal is exposed to additional post-processing, e.g. determining of  local maximums and  other peculiarities. In this study,  we avoid spectra post-processing and compared raw spectra, defined the frequency band of significant difference.

\section{Duality of Interest}
No potential conflicts of interest relevant to this article were reported.

\section*{Acknowledgment}

This work was supported by the grant of the President of the Russian Federation for state support of young Russian scientists № MK-7168.2016.8, RFBR-ra (project 17-41-590560). E Zherebtsov kindly acknowledges for personal support the funding from the European Union’s Horizon 2020 Research and Innovation Program under grant agreement No 703145.

Authors thank Dr Victor V. Sidorov and Prof Alexander I. Krupatkin for useful discussion. Special thanks are extended to doctors Alimicheva E.A., Masalygina G.I. and Muradyan V.F. of the “Orel Regional Clinical Hospital” for providing useful advice and help.


\end{document}